\documentclass[11pt,british,notitlepage]{article}
\usepackage[a4paper, left=2cm, right=2cm, top=2cm, bottom=2cm]{geometry}
\usepackage[numbers,sort&compress]{natbib}
\usepackage{amsmath, amssymb, amstext}
\usepackage{graphicx}
\graphicspath{ {Plots/} }
\usepackage{ucs}
\usepackage[utf8x]{inputenc}
\usepackage[T1]{fontenc}
\usepackage[DIV=15]{typearea}
\usepackage{subcaption}
\usepackage{lettrine}
\usepackage{float}
\usepackage{tabularx}
\usepackage{multirow}
\usepackage{xcolor}

\begin{document}

%%%%%%%%%%%%%%%%%%%%%%%%%%%%%%%%%%%
% TITLE
%%%%%%%%%%%%%%%%%%%%%%%%%%%%%%%%%%%

\title{\bf {Collisional alignment and molecular rotation control chemi-ionization of individual conformers}}
\author{L. Ploenes$^{1\ddag}$, P. Stra\v{n}\'{a}k$^{1,a\ddag}$, A. Mishra$^1$, X. Liu$^2$, J. P\'erez-R\'ios$^{3,4*}$ and S. Willitsch$^{1*}$}
\date{}
\maketitle

\begin{center}
$^1$ Department of Chemistry, University of Basel, Klingelbergstrasse 80, 4056 Basel, Switzerland \\
$^2$ Fritz-Haber-Institut der Max-Planck- Gesellschaft, D-14195 Berlin, Germany \\
$^3$ Department of Physics and Astronomy, Stony Brook University, Stony Brook, New York 11794, USA \\
$^4$ Institute for Advanced Computational Science, Stony Brook University, Stony Brook, New York 11794, USA \\
$^a$ Present address: Fraunhofer Institute for Applied Solid State Physics IAF, Tullastrasse 72, 79108 Freiburg, Germany \\
$\ddag$ These authors contributed equally to this work\\
$\ast$ Electronic mail: stefan.willitsch@unibas.ch, jesus.perezrios@stonybrook.edu
\end{center}

%%%%%%%%%%%%%%%%%%%%%%%%%%%%%%%%%%%
% ABSTRACT
%%%%%%%%%%%%%%%%%%%%%%%%%%%%%%%%%%%

{\bfseries\noindent

The relationship between the shape of a molecule and its chemical reactivity is a central tenet in chemistry. However, the influence of the molecular geometry on reactivity can be subtle and result from several opposing effects. Using a novel crossed-molecular-beam experiment in which individual rotational quantum states of specific conformers of a molecule are separated, we study the chemi-ionization reaction of hydroquinone with metastable neon atoms. We show that collision-induced alignment of the reaction partners caused by geometry-dependent long-range forces crucially influences reaction pathways for distinct conformers which is, however, countered by molecular rotation. We demonstrate how the interplay between molecular geometry, chemical steering forces and rotational dynamics govern the outcome of reactions and illustrate the capability of advanced molecule-control techniques to unravel these effects.
}

\bigskip

%%%%%%%%%%%%%%%%%%%%%%%%%%%%%%%%%%%
% INTRODUCTION
%%%%%%%%%%%%%%%%%%%%%%%%%%%%%%%%%%%
 
The geometry of molecules affects their physical properties, interactions and chemical functions \cite{eliel94a,frauenfelder91a,robertsona01a, kim07a, champenois21a}. In larger molecules, conformers, i.e., versions which convert into one another through rotations about chemical bonds, represent the most abundant form of isomers. Their thermal interconversion under ambient conditions imposes great challenges for unravelling their distinct reactivities. Only recently, experimental advances have enabled conformationally specific reaction studies under single-collision conditions in the gas phase yielding valuable insights into  reaction mechanisms of individual conformers \cite{taatjes13a,lin15a,chang13a,kilaj21a}.

Conformational effects are often steric in character and thus intertwined with stereodynamics implying a specific orientation of the reaction partners during collisions. Orientation is affected by geometry-dependent intermolecular interactions, but may also be influenced by rotational motion of the molecules which counters any orienting forces. The reactivity of a molecule will thus be influenced by the interplay of these effects. 

Here, we focus on this interrelation between molecular conformation, stereo- and rotational dynamics in chemical reactions. We deployed a novel experiment utilising electrostatic deflection to separate individual conformers of 1,4-dihydroxybenzene (hydroquinone, HYQ) within a molecular beam. These are intersected with another beam containing metastable neon atoms in the excited (2p)$^5$(3s)$^1$~\textsuperscript{3}P\textsubscript{2,0} states (Ne$^*$) \cite{ploenes21a} which ionize the molecular collision partner. Such chemi-ionization (CI) reactions are important in energetic environments like plasmas, planetary atmospheres and interstellar space \cite{siska93a, gordon20a} and have recently been used to explore quantum effects in reactive collisions \cite{henson12a, margulis23a}. Under hyperthermal conditions, CI occurs predominantly via an electron-exchange mechanism contingent on the overlap between orbitals of the molecule and the metastable. These processes are thus strongly affected by stereodynamic factors \cite{ohoyama99a, kishimoto07a, ascenzi19a, falcinelli20a, falcinelli20b, shagam15a, klein17a, gordon18a, zou19a, gordon20a} and represent an ideal testbed for the present investigation.

\begin{figure}[!h]
\centering
\includegraphics[width=\textwidth]{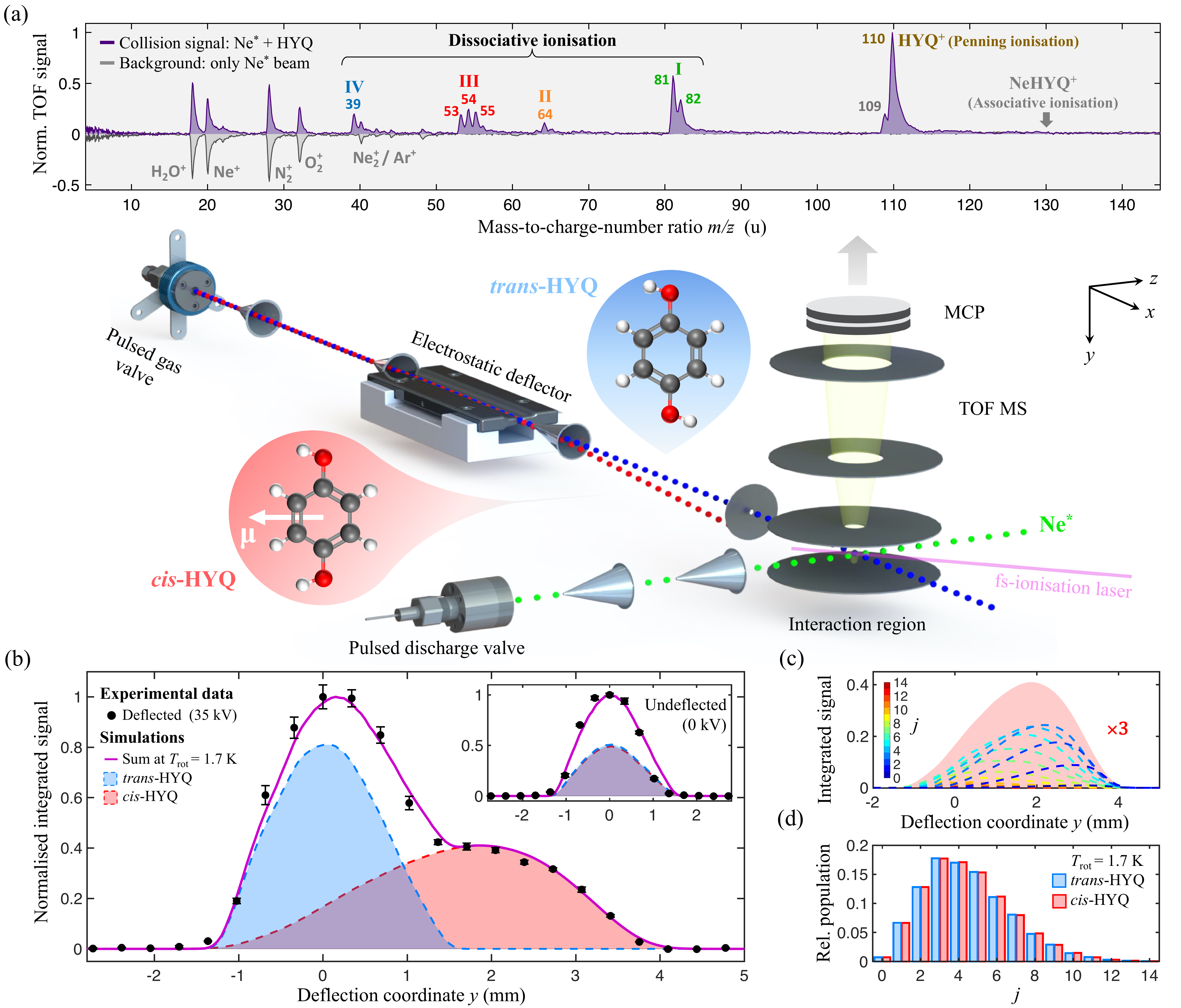}
\caption{Experiment and conformer separation by electrostatic deflection. (a) Schematic of the setup and geometries of the polar \emph{cis-} and apolar \emph{trans-}hydroquinone (HYQ). Inset: time-of-flight mass spectrum of the products of the chemi-ionization reaction of HYQ with Ne$^*$. %Intensities are normalized with respect to the strongest signal. 
Lower inverted trace: background experiment without the HYQ beam. (b) Density profile of the deflected HYQ beam (data points). Shaded areas: Monte-Carlo trajectory simulation of the deflection profile of the \textit{trans}- (blue) and \textit{cis}- (red) conformers and their sum (magenta trace). Inset: same data for an undeflected HYQ beam. Error bars represent the standard error of three individual measurements with each data point averaged over 2000 experimental cycles. (c) Simulated deflection profiles for specific rotational states $j$ and their sum (red-shaded area) for \emph{cis}-HYQ. (d) Relative populations of rotational levels at a temperature $T_{\text{rot}}=1.7$~K.}
\label{fig:Figure1}
\end{figure}

HYQ is an attractive model system as its CI reaction with Ne\textsuperscript{*} proceeds via different pathways resulting in Penning-ionization~(HYQ\textsuperscript{+}), associative-ionization~(NeHYQ\textsuperscript{+}) and several dissociative-ionization products. The rotation of the hydroxyl groups around the bonds with the benzene ring is inhibited leading to an apolar \textit{trans}- and a polar \textit{cis}-conformer (Figure~\ref{fig:Figure1}~(a)). The different dipole moments lead to markedly different interactions with Ne$^*$ and also enable a spatial separation of the conformers using electrostatic deflection \cite{chang15a, kilaj18a, kilaj20a, you18a}. 

%%%%%%%%%%%%%%%%%%%%%%%%%%%%%%%%%%%
% RESULTS
%%%%%%%%%%%%%%%%%%%%%%%%%%%%%%%%%%%

\section*{Experimental approach}

Using a novel crossed-molecular-beam apparatus \cite{ploenes21a}, we introduce here a new approach to study conformational effects in reactions of neutral molecules following our previous work on ion-molecule processes \cite{chang13a, kilaj21a}. A schematic of the setup is presented in Figure~\ref{fig:Figure1}~(a), see Methods for details. Briefly, a molecular beam containing a mixture of the two HYQ conformers was injected into the inhomogeneous electric field of an electrostatic deflector where the molecules were spatially separated according to their effective dipole moments. Density profiles were recorded by tilting the molecular-beam machine and thus overlapping different regions of the deflected HYQ beam with a femtosecond (fs) laser beam ionising the molecules. The ions were analysed by a time-of-flight mass spectrometer (TOF-MS). For the reaction experiments, the laser was replaced by a Ne$^*$ beam.  

%\section*{Results}

\section*{Spatial separation of \textit{trans}- and \textit{cis}-hydroquinone in different rotational states}\label{sec:HYQDeflExpCH5}

A density profile of the deflected HYQ beam is shown in Figure~\ref{fig:Figure2}~(b). The data were obtained by integrating the TOF mass signal of HYQ\textsuperscript{+} for different tilting angles of the molecular-beam translated into a deflection coordinate $y$ representing the offset from the center of the undeflected beam. 

Comparing Monte-Carlo trajectory simulations of molecules propagating through the deflection setup \cite{chang13a, kilaj18a, kilaj20a} with the experiment allowed the determination of the individual deflection profiles for each conformer \cite{kilaj20a} (blue and red areas in Figure~\ref{fig:Figure1}~(b)) revealing the separation of the deflected \textit{cis}- from the apolar and hence undeflected \textit{trans}-conformer (see also Ref.~\cite{you18a}). 

The degree of deflection depends on the effective dipole moment of the HYQ molecules which is also influenced by their rotational state \cite{chang15a, kilaj18a, kilaj20a, ploenes21a}. For \textit{cis}-HYQ, this is illustrated in Figure~\ref{fig:Figure1}~(c) which shows deflection profiles for the experimentally populated rotational states (Figure~\ref{fig:Figure1}~(d)) with quantum numbers $j$ and summed over all asymmetric-top quantum numbers~$\tau$. Molecules in low rotational states had a larger effective dipole moment at the experimentally relevant field strengths and were deflected more strongly. 

\section*{Chemi-ionization of Ne\textsuperscript{*} with conformationally selected hydroquinone}\label{sec:HYQCINeResults}

A TOF mass spectrum of the reaction products is shown in Figure~\ref{fig:Figure1}~(a). Besides the PI product HYQ\textsuperscript{+} at a mass-to-charge-number ratio $m/z=110$, various ionic reaction products resulting from DI (labelled I-IV) can be observed. An associative-ionization product NeHYQ\textsuperscript{+} was not detected. The DI fragment ions were identified based on their $m/z$ ratios and earlier studies \cite{akopyan00a,hassan16a,holstein01a}. Their proposed chemical formulae are summarized in Table~\ref{tab:HYQReacProducts} together with the product branching ratios. The branching ratio of the PI pathway to the sum of all DI pathways was determined to be $\text{PI}:\text{DI}=1:1.2(1)$.  

\begin{table}[!h]
\begin{tabularx}{\textwidth} {>{\raggedright\arraybackslash}l>{\centering\arraybackslash}X>{\centering\arraybackslash}c>{\raggedleft\arraybackslash}c}
\hline
\hline
\textbf{Label}           & \textbf{Reaction products} & \textbf{Mass of ionic product (u)} & \textbf{Branching ratio}   \\ \hline \hline
PI      &  C$_6$H$_6$O$_2^+$ & 110 & 1 \\ \hline
\multirow{2}{*}{I}       & C$_5$H$_6$O$^+$ + CO & 82 & \multirow{2}{*}{0.61(4)}   \\ 
        & C$_5$H$_5$O$^+$ + HCO & 81 & \\ \hline
II       & C$_5$H$_4^+$ + CO + H$_2$O & 64 & 0.04(1)  \\  \hline
\multirow{3}{*}{III}       & C$_3$H$_3$O$^+$ + C$_3$H$_3$O & 55 & \multirow{3}{*}{0.43(3)}  \\ 
        & C$_3$H$_2$O$^+$ + C$_3$H$_4$O & 54 &  \\
       & C$_3$HO$^+$ + C$_3$H$_5$O & 53 & \\ \hline
IV       & C$_3$H$_3^+$ + C$_2$H$_3$O + CO & 39 & 0.12(1)  \\
\hline \hline
\end{tabularx}
\caption{Products and branching ratios of the chemi-ionization reaction of Ne\textsuperscript{*} + HYQ. Error bars represent the standard error of 6 individual measurements averaged over 10 000 experimental cycles. }
\label{tab:HYQReacProducts}
\end{table}

\begin{figure}[!b]
\centering
\includegraphics[width=\textwidth]{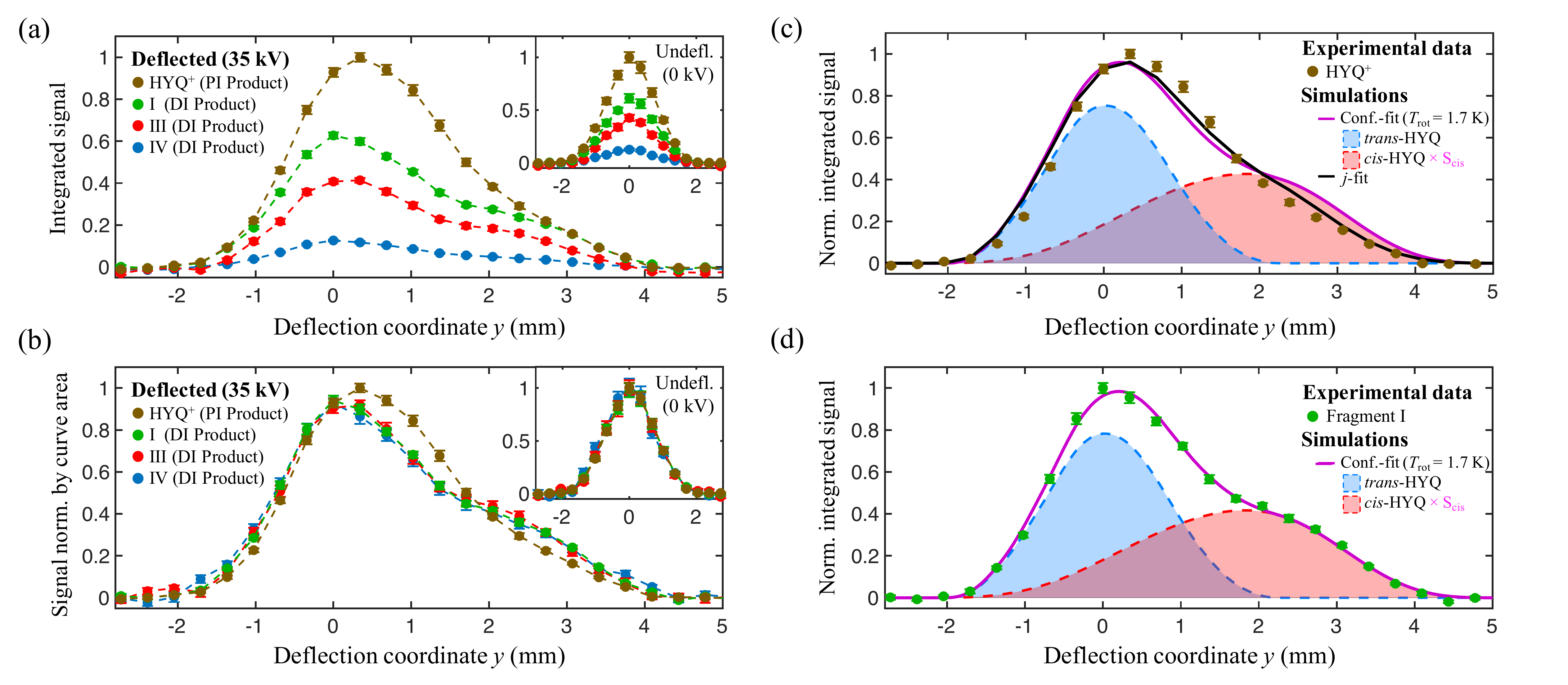}
\caption{CI reaction-deflection profiles of HYQ + Ne\textsuperscript{*}. (a) Comparison of profiles for the PI product HYQ\textsuperscript{+} and different DI products for a deflected molecular beam. Inset: same data for an undeflected molecular beam. Error bars represent the standard error of 18 individual traces with each data point averaged over 10 000 experimental cycles. (b) Same deflection profiles additionally normalized by their area. (c, d) Least-square fits of conformationally-specific simulated reaction-deflection profiles to the experimental data for the (c) PI and (d) DI reaction pathways.}
\label{fig:Figure2} 
\end{figure}

Figure~\ref{fig:Figure2}~(a) shows reaction-deflection profiles obtained by integrating the relevant mass peaks in TOF spectra at different deflection coordinates $y$ for the PI product HYQ\textsuperscript{+} as well as the DI products I, III and IV (Table~\ref{tab:HYQReacProducts}). The profiles were normalized with respect to HYQ$^+$. Figure~\ref{fig:Figure2}~(b) depicts the same profiles additionally normalized by their areas to correct for the product branching ratios. Strikingly, the profiles of all DI products in Figure~\ref{fig:Figure2}~(b) overlap, while the PI-product profile is clearly different. This observation contrasts with the data obtained for an undeflected HYQ beam, in which all profiles independent of the reaction pathway neatly overlap (inset of Figure~\ref{fig:Figure2}~(b)). We thus conclude that distinct specificities must exist for the PI and DI pathways. Because the profiles of all DI products appear to be identical within the experimental uncertainties, only product channel I will be considered henceforth.  

To analyse these results, least-square fits of simulated deflection profiles to the experimental data were performed. First, the total contribution of \textit{trans}- and \textit{cis}-HYQ to the reaction-deflection profiles was analysed to capture \emph{purely conformational} differences assuming the reactivity to be equal for all rotational states. Conformationally specific reactivities were expressed by a fit parameter $S_{\text{cis}}$ which scales the profile of the \textit{cis}-conformer with respect to \textit{trans}-HYQ relative to their initial populations in the beam. Results are displayed as the magenta traces in Figure~\ref{fig:Figure2}~(c, d).

For both reaction pathways, the fits yielded only insignificant differences between the conformers with \textit{cis}-HYQ showing around 8~\% and 6~\% reduced reactivity compared to the \textit{trans}-species for the PI and DI channels, respectively. The experimental data for the DI products (exemplified by product I in Figure \ref{fig:Figure2} (d)) is nicely reproduced by the trajectory simulations with $S_{\text{cis}}\approx0.94(1)$. By contrast, the experimental results for the PI pathway could not be well replicated by the simulations when only conformational reactivities were adjusted (Figure~\ref{fig:Figure2}~(c) with $S_{\text{cis}}\approx0.92(5)$). 

The differences between experiment and the magenta trace in Figure~\ref{fig:Figure2}~(c) suggest that \emph{cis-}HYQ in low rotational states, which are most strongly deflected and thus appear at larger deflection coordinates, show a decreased reactivity in PI compared to the simulation, whereas no evidence for such a propensity was observed for DI (Figure~\ref{fig:Figure2}~(d)). Consequently, the black trace in Figure~\ref{fig:Figure2} (c) shows a simulation in which the individual reactivities of the different $j$ states in the PI channel were fitted to the experimental data. The fit yielded negligible contributions of the lowest rotational states to the reaction-deflection profile (see Supplementary Information and Figure~\ref{fig:FigureS1} for details) resulting in a markedly improved agreement between simulation and experiment.  

%%%%%%%%%%%%%%%%%%%%%%%%%%%%%%%%%%%
% DISCUSSION
%%%%%%%%%%%%%%%%%%%%%%%%%%%%%%%%%%%

%\clearpage
\section*{Reaction dynamics}

These results demonstrate that for \emph{cis-}HYQ, the relative reactivities towards PI and DI, and thus the product-branching ratios, depend on the rotational state. The mechanisms for CI reactions have been discussed previously, see, e.g., References~\cite{siska93a, kishimoto07a, falcinelli20a}. They typically proceed via an exchange mechanism in which an electron from a molecular orbital (MO) of the reagent is transferred to the singly occupied valence orbital of the metastable (here a 2p~orbital of Ne$^*$) followed by the ejection of the excited electron. The efficiency of this process is governed by the overlap of the singly occupied valence orbital of the metastable and the relevant molecular orbital of its collision partner at the distance of closest approach around the turning point of the collision.

\begin{figure}[!t]
\begin{minipage}{0.6\textwidth}
        \centering
        \includegraphics[width=\textwidth]{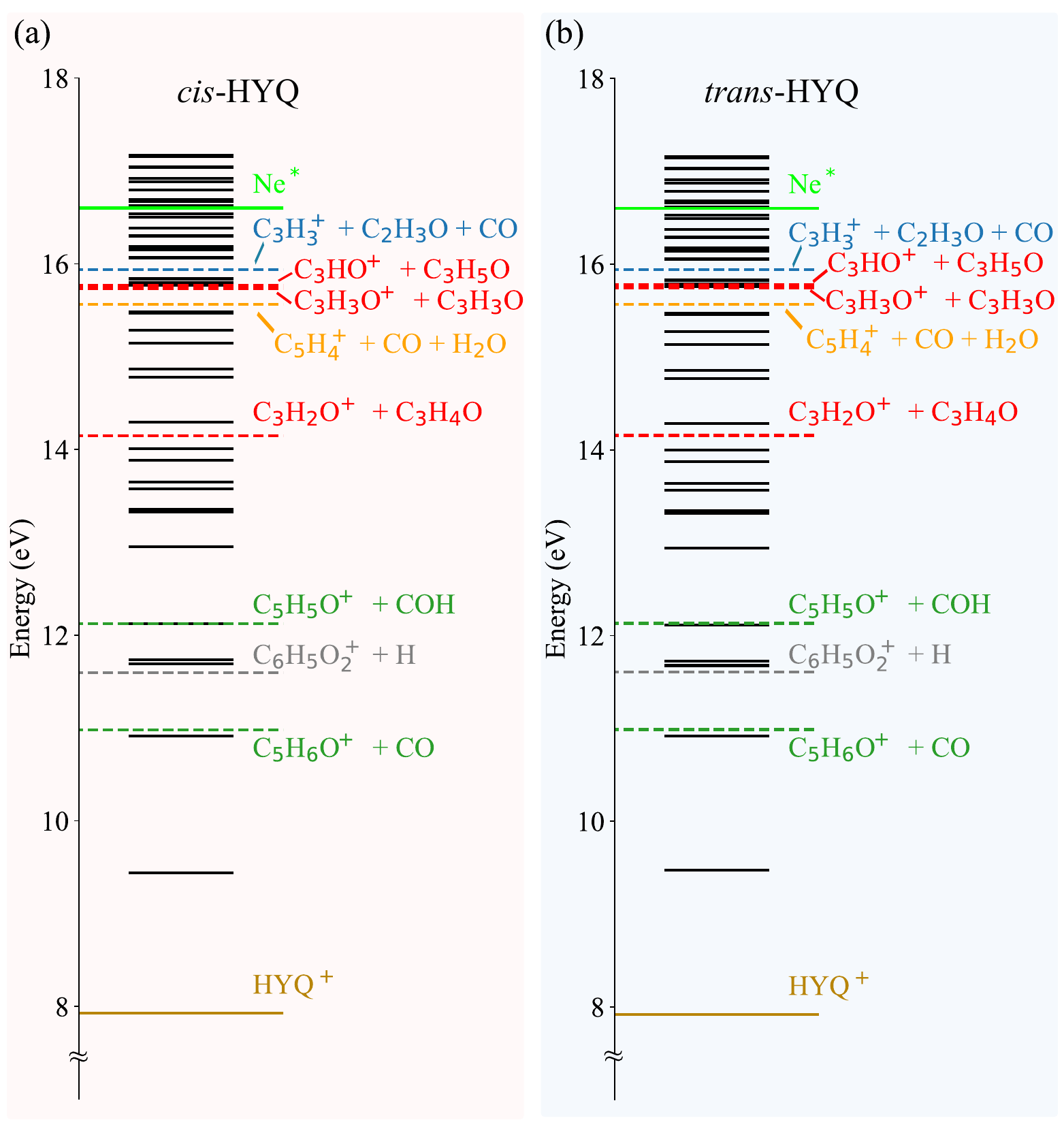}
\end{minipage}
\hfill
\begin{minipage}{0.38\textwidth}
        \caption{Energetics of CI processes. Calculated energies of electronic states and dissociation thresholds of the (a) \emph{cis}- and (b) \emph{trans}-HYQ$^+$ ion referenced to the neutral ground state S$_0$.}
        \label{fig:Figure3}
\end{minipage}
\end{figure}
    
Supplementary Figures~\ref{fig:FigureS2} (a) and (b) show the highest occupied and lowest unoccupied MOs of \emph{cis}- and \emph{trans}-HYQ, respectively. CI is initiated by the removal of an electron from an occupied MO in the collision complex with Ne$^*$, generating HYQ$^+$ in a specific electronic state which may further dissociate to DI products. Calculated energies of the lowest electronic states of \emph{cis}- and \emph{trans}-HYQ$^+$ are displayed in Figure~\ref{fig:Figure3}~(a) and (b), respectively. Dissociation thresholds of the DI products in Table~\ref{tab:HYQReacProducts} are indicated by dashed lines (see Supplementary Figure~\ref{fig:FigureS3} for their structures). It can be surmised that the PI product HYQ$^+$ originates from the formation of the lowest three ionic states D$_0$, D$_1$ and D$_2$ below the lowest dissociation limit (leading to C$_5$H$_6$O$^+$ + CO), while the population of higher-lying electronic states results in DI. 

Inspection of the electronic structure of the lowest three ionic states in Figure~\ref{fig:FigureS2}~(c, d) reveals that the $D_0$~ground state mainly originates from the removal of a single electron from the highest occupied molecular orbital (HOMO) of HYQ. $D_1$ and $D_2$ involve additional excitations from lower-lying MOs such that their electronic configuration differs by more than one electron from the $S_0$ neutral ground state. The transition moments for such multi-electron rearrangements can be expected to be small in comparison to the one-electron process leading to $D_0$. Thus, only the generation of $D_0$ appreciably contributes to the observed PI process and the efficiency of PI is then governed by the overlap of the 2p valence orbitals of Ne$^*$ with the HOMO of HYQ during the collision.

\begin{figure}[!h]
        \centering
        \includegraphics[width=1\linewidth]{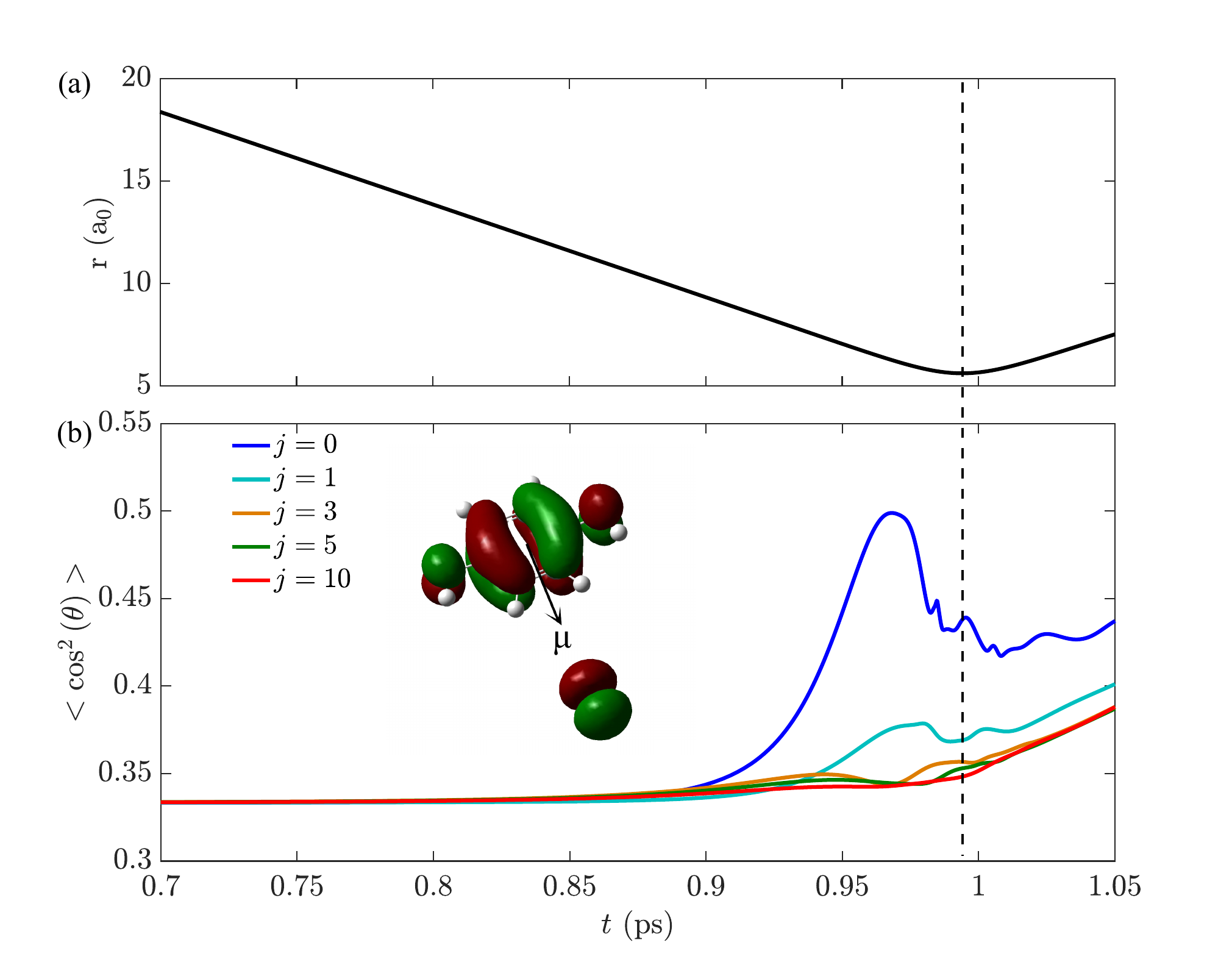}
        \caption{Collision-induced alignment. (a) Trajectory (interparticle distance~$r$ as function of time~$t$) for a collision of Ne$^*$ and HYQ. (b) Alignment $\langle\cos^2(\theta)\rangle$ as a function of time for different initial rotational states~$j$ of HYQ. Inset: illustration of the alignment along the axis of the molecular dipole~$\mu$ during the collision.}
        \label{fig:Figure4}
    \end{figure}

Based on these findings, the rotational dependence of PI of the \emph{cis}-conformer can be rationalized by a stereodynamic effect. During the attack of Ne$^*$ on \emph{cis}-HYQ, the molecule aligns towards the metastable by the interaction of its permanent dipole with a dipole induced in the highly polarizable Ne$^*$. The alignment steers the direction of attack along the axis of the dipole moment~$\mu$ in the nodal plane of the HOMO. For an attack trajectory close to the dipole axis  (see illustration in inset of Figure~\ref{fig:Figure4}~(b)), the overlap between the HOMO and any of the 2p~orbitals of Ne$^*$ is small because of cancellation effects resulting in a poor efficiency for PI. However, this alignment is hindered by the rotation of HYQ: It is most effective in low rotational states which exhibit the largest effective dipole moments and it is suppressed with increasing rotational excitation. It can thus be expected that low rotational states of HYQ show a decreased tendency towards PI, in line with the experiments.

While a rigorous \emph{ab-initio} treatment of the CI reaction of a system of the present size is currently beyond reach, the salient features of the dynamics can be qualitatively captured using a simple model. Figure~\ref{fig:Figure4} shows the results of a quasiclassical trajectory (QCT) calculation which assumed a classical collision of a polarizable particle (Ne$^*$) with a dipole treated as a quantum rotor (HYQ) at zero impact parameter. The resulting dipole-induced dipole-interaction potential is displayed in Supplementary Figure~\ref{fig:FigureS4}. The translational degrees of freedom were treated classically leading to an effective time-dependent potential acting as alignment field resembling the mixed quantum-classical approach of Reference~\cite{Ivanov2011}. The time-dependent Schr\"odinger equation for the rigid rotor was then solved under the action of this time-dependent potential (see Methods). 

Figure~\ref{fig:Figure4}~(a) shows a typical collision trajectory at an energy of 0.5~eV. Figure~\ref{fig:Figure4}~(b) illustrates the alignment $\langle \cos^2(\theta)\rangle$ along the trajectory for different initial rotational states $j$ of HYQ, where $\theta$ is the angle between the molecular dipole and the collision axis and $\langle..\rangle$ denotes an expectation value. For any initial rotational state, the alignment is 0.33 at long range characteristic of a randomly oriented sample. The alignment increases markedly upon approach of the collision partners due to the dipole-induced dipole interaction. Because of the high collision energy in the present experiment ($\approx0.5$~eV) and the short-range nature of the potential scaling as $V(r)\sim r^{-6}$ with distance $r$, the alignment is essentially impulsive (see Supplementary Figure~\ref{fig:FigureS5}). The electron transfer giving rise to CI is expected to occur around the turning point of the trajectory (dashed line in Figure~\ref{fig:Figure4}) where the orbital overlap is maximal and the collision velocity minimal. Figure~\ref{fig:Figure4}~(b) illustrates that the alignment at the turning point decreases sharply with increasing quantum number~$j$ of the rotor: It is maximal for $j=0$ and essentially negligible for $j>5$. Note that the most populated rotational state in the experiment is $j=3$ and about 40\% of the population is contained in levels with $j\leq 3$ (Figure \ref{fig:Figure1} (d)). 

While this simple model neglects details like the exact short-range potential of the collision, the different channels arising from the $^3P_{2,0}$ states of Ne$^*$, the asymmetry of the HYQ quantum rotor and detailed assumptions on the CI decay widths, it captures the important stereodynamics in the entrance channel of the reaction and thus provides a qualitative explanation of the experimental results. We confirmed numerically that the calculated alignment indeed leads to a $j$-dependence of the relevant overlap integrals implying a reduced reactivity in PI for the lowest $j$ states.

The alignment dynamics of \emph{cis-}HYQ are of course also expected to be active for the DI channels. However, any stereodynamic effects arising from the specific shape of the molecular orbitals involved are anticipated to be averaged out by the multitude of cationic electronic states, and therefore molecular orbitals, which contribute to the DI products in the present experiment. These are all the states located between the first dissociation channel of HYQ$^+$ and the energy of Ne$^*$ indicated in Figure~\ref{fig:Figure3}. Indeed, the DI channels observed for \emph{cis}-HYQ do not exhibit any noticeable rotational dependencies as discussed above. 

From the available experimental data, no conclusions can be drawn on \emph{trans}-HYQ which is not deflected in our experiments due to the lack of a dipole moment. However, it can be conjectured that any stereodynamic effects due to alignment during the collision are significantly weaker than in the \emph{cis} species because of the absence of long-range dipolar interactions.

%%%%%%%%%%%%%%%%%%%%%%%%%%%%%%%%%%%
% CONCLUSIONS
%%%%%%%%%%%%%%%%%%%%%%%%%%%%%%%%%%%

\section*{Conclusions}

In the present study, we have explored the interrelation between molecular geometry, chemical steering forces and molecular rotation governing the dynamics of the CI reaction of individual conformers of hydroquinone with metastable neon. The present findings on the CI of \emph{cis}-HYQ can be contrasted with the rotational-state dependencies \cite{kilaj18a, li14b} as well as alignment dynamics \cite{cernuto17a, cernuto18a} recently uncovered in ion-molecule reactions. Whereas those processes were controlled by strong ion-dipole long-range interactions, the dipole-induced dipole couplings dominating HYQ + Ne$^*$ represent by comparison an interaction regime of intermediate strength in which intermolecular forces are strong enough to generate significant alignment effects, but molecular rotation is still able to moderate them. Moreover, these interactions and thus this dynamical motif are strongly dependent on the molecular conformation, as shown here. Our results reveal a dynamical pattern which can be expected to be common in chemical dynamics of reactions such as CI which are governed by strong steric or stereoelectronic effects. The present study also illustrates the power of advanced techniques to control the motion and quantum states of molecules in the gas phase for the elucidation of the chemistry of complex molecules.

%%%%%%%%%%%%%%%%%%%%%%%%%%%%%%%%%%%
% REFERENCES
%%%%%%%%%%%%%%%%%%%%%%%%%%%%%%%%%%%

%\bibliography{MainBibFile, SupBibFile}
%\bibliographystyle{Science}

%%%%%%%%%%%%%%%%%%%%%%%%%%%%%%%%%%%
% METHODS
%%%%%%%%%%%%%%%%%%%%%%%%%%%%%%%%%%%

%\clearpage\pagebreak

\section*{Methods}
\label{sec:methods}

\subsection*{Experimental methods}

The principle of electrostatic deflection for the spatial separation of rotational states and conformers depending on their effective dipole moments was described in previous publications, see, e.g., \cite{filsinger09a, chang13a, chang15a, kilaj18a, kilaj20a}. These publications also contain a detailed explanation of the Monte-Carlo simulations of molecule trajectories through the deflection setup which were used in the analysis of the experimental data. The spatial separation of HYQ by electrostatic deflection was previously demonstrated in Reference~\cite{you18a}. In the present study, a molecular beam was generated from commercial HYQ (Sigma Aldrich, >~99~\%) heated to 150~\textdegree C in a pulsed gas valve from which the vapour pressure seeded in 90~bar of helium was expanded into high vacuum. The resulting molecular beam contained both rotamers of HYQ. As the energy difference between the two conformers is only on the order of 10~cm$^{-1}$ \cite{caminati94a}, they populated the beam in an approximately equal ratio of 51:49 (\textit{trans}:\textit{cis}) at the valve temperature. The beam containing both HYQ conformers was guided through the inhomogeneous electric field of an electrostatic deflector \cite{chang15a, chang13a} at a deflector voltage of 35~kV, which resulted in a spatial separation of the apolar \emph{trans}- and individual rotational states of  the polar \emph{cis}-conformer (dipole moment $\mu$ = 2.38~D) as shown schematically in Figure~\ref{fig:Figure1} (a). 

The density profile of different parts of the deflected molecular beam was probed in the interaction region by multi-photon ionization using the radiation of a near-infrared femtosecond laser (Clark-MXR CPA 2110, 775~nm) followed by time-of-flight mass spectrometry (TOF-MS). Thus, "deflection profiles", i.e., density profiles of the  molecular beam as a function of the deflection along the coordinate $y$, were obtained. Comparison with Monte-Carlo trajectory simulations enabled the characterization of the conformational and rotational-state composition of molecules at different positions of the beam as well as the determination of their rotational temperature $T_{\text{rot}}=1.7$~K (see Refs. \cite{ploenes21a, kilaj18a, kilaj20a} for details of the procedure).

For studying CI reactions, the deflected beam of HYQ molecules was crossed with a beam of metastable neon atoms in the (2p)$^5$(3s)$^1$~\textsuperscript{3}P\textsubscript{2} and \textsuperscript{3}P\textsubscript{0} states generated in a supersonic expansion of neon gas (stagnation pressure 45~bar) through a pulsed plate-discharge source \cite{stranak21a}. Different parts of the deflected beam, containing unique compositions of HYQ rotamers in different rotational states, were overlapped with the Ne$^*$ beam in the collision region by tilting the HYQ molecular-beam setup with respect to the collision chamber \cite{ploenes21a}. The resulting ionic reaction products were analysed using TOF mass spectrometry. This approach enabled the investigation of conformer- and state-specific reactivities towards the different reaction pathways of the reaction of HYQ with metastable neon. From the beam velocities, the collision energy of the reaction was determined to be approximately 0.5~eV.

Several signals in the TOF mass spectrum shown in the inset of Figure~\ref{fig:Figure1} (a) (H\textsubscript{2}O\textsuperscript{+}, N\textsubscript{2}\textsuperscript{+}, O\textsubscript{2}\textsuperscript{+}) could be assigned to PI of trace gases in the chamber by comparison with a spectrum recorded without the HYQ beam (grey inverted trace). Features corresponding to Ne\textsuperscript{+} and Ne$_2^+$ could be attributed to intra-beam PI and AI between two metastable neon atoms, respectively.

The product branching ratios listed in Table~\ref{tab:HYQReacProducts} were determined by integrating the relevant signals in the TOF mass spectrum and comparing them to the one of the PI product. Note that also a small signal at $m/z=109$ can be observed in the mass spectrum in \ref{fig:Figure1} (a) which originates from the dissociation of a single hydrogen atom from HYQ$^+$. As this peak is not sufficiently resolved from the much stronger HYQ\textsuperscript{+} signal, it was not included in the analysis of the product branching ratios.

In Figures~\ref{fig:Figure2} (a) and (b), the deflection profile of the weak signal associated with DI product II in the TOF-MS was not included because of poor signal-to-noise ratio, but follows the same trend as the other DI products within the uncertainty limits.

\subsection*{Theoretical methods}

\subsubsection*{Quantum chemistry calculations}
The title reaction was studied using quantum chemistry methods. The ground states of the reactants and products of the present CI reaction have been calculated with density functional theory (DFT) using the hybrid exchange–correlation functional CAM-B3LYP \cite{yanai2004new} as implemented in Gaussian16~\cite{g16}. The energy landscape involved in the reaction was calculated with time-dependent density functional theory (TDDFT) using the same functional as for the ground state but considering 400 excited states up to an excitation energy of 17~eV, approximately covering the available reaction energy (internal energy of Ne$^*$ plus the collision energy). The def2-QZV basis set \cite{weigend2005balanced} has been used for neutral and ionic HYQ and their excited states. For Ne, the aug-cc-pVTZ basis set \cite{ccdunning1989gaussian, cckendall1992electron} has been employed. The same approach has been adopted for both the \textit{cis-} and \textit{trans-}conformer of hydroquinone. The energies of excited states in Figure~\ref{fig:FigureS2}~(c, d) of the Supplementary Information were obtained from the computationally optimized geometries, and referenced to the energy of the optimized S$_0$~ground state of HYQ. 

\subsubsection*{Quasi-classical trajectory calculations}

The alignment of HYQ in collisions with Ne$^*$ under the action of their dipole-induced dipole interaction was modelled using quasiclassical trajectory calculations. The following model potential was adopted:
 \begin{equation}
\label{eqqc1}
V(r,\theta)=\frac{C_{12}}{r^{12}}-\frac{\alpha \mu^2}{2r^6}\left(1+3\cos^2(\theta) \right),
\end{equation}
where $\mu=2.38$~Debye \cite{caminati94a} is the permanent dipole moment of the molecule, $\alpha=27.8$~\AA$^3$ is the polarizability of Ne\textsuperscript{*} in the \textsuperscript{3}P\textsubscript{2}~state~\cite{brunetti13a}, $r$ is the distance between the centre of mass of the molecule and the atom, and $\theta$ is the angle of the molecule with respect to $\vec{r}$. The first term in Equation~(\ref{eqqc1}) accounts for the typical repulsion when the electronic clouds of the atom and molecule overlap. We assumed a value $C_{12}=2.5\times 10^{7}$~a.u. leading to a distance of closest approach similar to 5a$_0$. The potential is plotted in Figure~\ref{fig:FigureS4} of the Supplementary Information.

For collisions at zero impact parameter, the classical Hamiltonian of the system is given by

\begin{equation}
H=\frac{p_r^2}{2\mu_\text{c}}+\frac{2}{m_{\text{HYQ}}}p_\theta^2+ V(r,\theta),
\label{eqqc2}
\end{equation}
where $p_r$ and $p_\theta$ stand for the canonical momentum associated with $r$ and $\theta$, respectively, and \linebreak $\mu_\text{c}=(m_\text{Ne} m_{\text{HYQ}})/(m_\text{Ne}+m_{\text{HYQ}})$ is the reduced mass of the collision. The rotational dynamics is treated by approximating the molecule as a linear rotor with two point masses $m_{\text{HYQ}}/2$ situated at the extremities of a rigid rotor of length $R_e$, representing the equilibrium distance. By solving Hamilton's equations, using appropriate initial conditions within the quasiclassical trajectory method (see Supplementary Information for further details), it is possible to infer a time-dependent external potential for the rigid rotor of the form

\begin{equation}
\label{eqqc3}
V_\text{ext}(t,\theta)=\frac{C_{12}}{r^{12}}-\frac{\alpha \mu^2}{2r(t)^6}\left(1+3\cos^2(\theta) \right),
\end{equation}
which can be viewed as giving rise to an alignment field acting on the molecule. In this scenario, the time-dependent Schr\"odinger equation (TDSE) in atomic units reads as

\begin{equation}
\label{eqqc4}
\imath \frac{\partial}{\partial t}\Psi(\theta,\phi,t)=B_e\hat{j}^2\Psi(\theta,\phi,t)+V_{ext}(t, \theta)\Psi(\theta,\phi,t),
\end{equation}
in which the first term of the right-hand side of the equation represents the rotational energy of the molecule with angular-momentum operator $\hat{j}$, $\Psi$ is the wavefunction and $B_e$ the rotational constant of the linear rotor ($B_e=5.614$~GHz). Using the following ansatz

\begin{equation}
\label{eqqc5}
\Psi(\theta,\phi,t)=\sum_{j}C_{jm}(t)|jm\rangle e^{-\imath B_ej(j+1) t},
\end{equation}
where $|jm\rangle$ are the eigenfunctions of the rigid rotor, i.e., spherical harmonics $Y_j^{m}(\theta,\phi)$, Equation~(\ref{eqqc4}) reduces to a system of ordinary differential equations that is numerically solved (details are given in the Supplementary Information). Finally, the alignment is calculated as 

\begin{equation}
\label{eqqc6}
\langle \cos^2{\theta}\rangle=\langle \Psi(\theta,\phi,t)|\cos^2{(\theta)}|\Psi(\theta,\phi,t)\rangle=\sum_j\sum_{j'}C^{*}_{jm}(t)C_{j'm}(t)e^{-\imath Be \left( j'(j'+1)-j(j+1)\right)t}\langle j'm|\cos^2{(\theta)}|jm\rangle.
\end{equation}

%%%%%%%%%%%%%%%%%%%%%%%%%%%%%%%%%%%
% REST
%%%%%%%%%%%%%%%%%%%%%%%%%%%%%%%%%%%

\subsection*{Acknowledgements}
We thank Philipp Kn\"opfel, Grischa Martin, Georg Holderried and Anatoly Johnson (University of Basel) for technical support. We also thank Dr. Nabanita Deb for assistance with the experiments. We acknowledge helpful discussions with Profs. Sang Kyu Kim (KAIST), Edvardas Narevicius (Technical University of Dortmund), Malte Oppermann (University of Basel), Andreas Osterwalder (EPFL) and Fernando Pirani (University of Perugia). This work was supported by the Swiss National Science Foundation under grants no. BSCGI0\_157874, IZKSZ2\_188329, IZCOZ0\_189907 and by the University of Basel. J.P.-R. thanks the support of the Simons Foundation. 

\subsection*{Author contributions}
L.P. and P.S. acquired the experimental data. L.P. analysed the data. A.M. assisted with the experiments and performed a part of the fits. X.L. performed the quantum-chemical calculations. J.P-R. performed the QCT calculations. J.P.-R. and S.W. supervised the project. S.W. conceived the project. L.P., J.P.-R. and S.W. drafted the manuscript, all other authors contributed to the writing.

%%%%%%%%%%%%%%%%%%%%%%%%%%%%%%%%%%%
% SUPPLEMENTAL INFORMATION
%%%%%%%%%%%%%%%%%%%%%%%%%%%%%%%%%%%

% \clearpage\pagebreak

% \setcounter{page}{1}

% \vspace*{3cm}

% \begin{center}
% {\LARGE\bf 
% {Supplementary Information for \\ \bigskip\bigskip
% Collisional alignment and molecular rotation control chemi-ionization of individual conformers}}

% \bigskip \bigskip

% \author{L. Ploenes$^{1\ddag}$, P. Stra\v{n}\'{a}k$^{1,a\ddag}$, A. Mishra$^1$, X. Liu$^2$, J. P\'erez-R\'ios$^{3,4*}$ and S. Willitsch$^{1*}$}

% \bigskip\bigskip\bigskip\bigskip

% $^1$ Department of Chemistry, University of Basel, Klingelbergstrasse 80, 4056 Basel, Switzerland \\
% $^2$ Fritz-Haber-Institut der Max-Planck- Gesellschaft, D-14195 Berlin, Germany \\
% $^3$ Department of Physics and Astronomy, Stony Brook University, Stony Brook, New York 11794, USA \\
% $^4$ Institute for Advanced Computational Science, Stony Brook University, Stony Brook, New York 11794, USA \\
% $^a$ Present address: Fraunhofer Institute for Applied Solid State Physics IAF, Tullastrasse 72, 79108 Freiburg, Germany \\
% $\ddag$ These authors contributed equally to this work\\
% $\ast$ Electronic mail: stefan.willitsch@unibas.ch, jesus.perezrios@stonybrook.edu
% \end{center}

\clearpage\pagebreak

\section*{Supplementary Information}

\setcounter{figure}{0}
\renewcommand\thefigure{S\arabic{figure}}
\setcounter{subsection}{0}
\renewcommand\thesubsection{S\arabic{subsection}}

\subsection*{Fit of rotational-state-dependent reactivities}

The contributions of different rotational states of \emph{cis-}hydroquinone (HYQ) with quantum numbers $j$ to the total Penning-ionization (PI) reaction-deflection profile shown in Figure~\ref{fig:Figure2} (c) of the main text were determined in a fit of the individual simulated $j$-dependent deflection profiles (summed over all asymmetric-top components $\tau$) to the experimental data points. In this fit, deflection profiles of states in the range $0 \leq j \leq 10$ were included, where the relative weight of each profile was constrained within a factor ranging from 0 to 2 relative to the thermal population of the state in the molecular beam. As the apolar \emph{trans-}conformer was not deflected in the experiment, the measurement was not sensitive to possible $j$-dependent reactivities of this species. Consequently, for \emph{trans-}HYQ all rotational states were assumed to contribute to the reaction-deflection profile according to their thermal populations with an overall weight constrained to a factor between 0.8 and 1.2 relative to the total contribution of \emph{cis-}HYQ. 

\begin{figure}[!h]
        \centering
        \includegraphics[width=\textwidth]{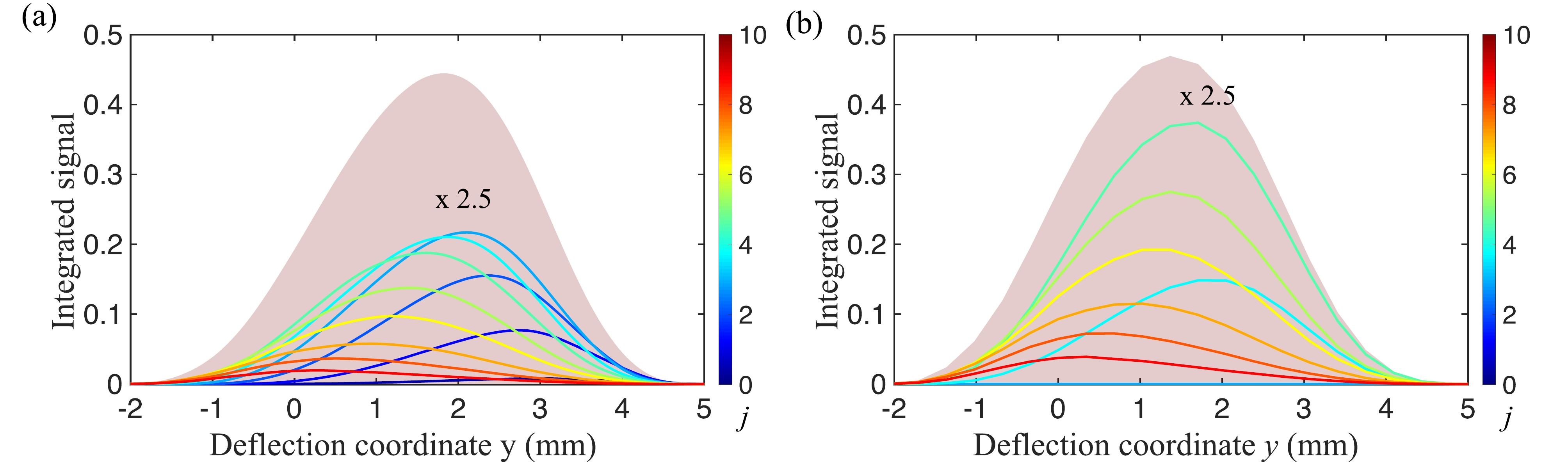}
        \caption{Decomposition of the simulated Penning-ionization reaction-deflection profiles for HYQ + Ne$^*$ shown in Figure~\ref{fig:Figure2} (c) of the main text into relative contributions of different rotational states with quantum number $j$. Only the contribution of the \emph{cis-}conformer is shown, the \emph{trans-}species has been omitted for clarity. The contributions of the different $j$ states to the reaction was (a) assumed to be proportional to their thermal population in the molecular beam (corresponding to the magenta trace in Figure~\ref{fig:Figure2} (c)), and (b) fitted to experimental data (corresponding to the black trace in Figure~\ref{fig:Figure2} (c)). The lowest rotational states were found to only negligibly contribute to the total reaction-deflection profile in (b).}
        \label{fig:FigureS1}
    \end{figure}

\begin{figure}[!h]
        \centering
        \includegraphics[width=\textwidth]{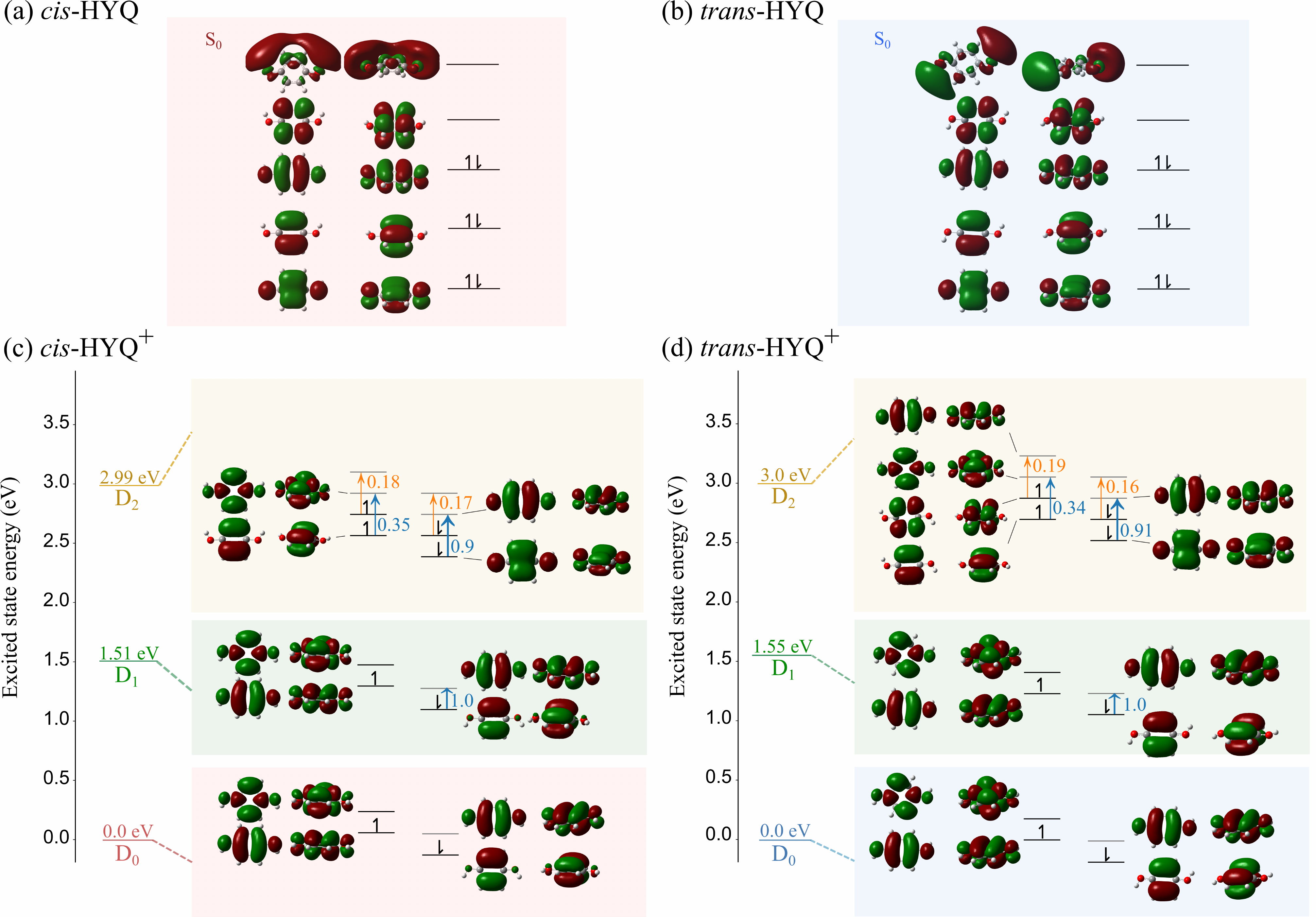}
        \caption{Molecular orbitals and electronic states of HYQ and HYQ$^+$. Highest occupied and lowest unoccupied molecular orbitals of (a) \emph{cis}- and (b) \emph{trans}-HYQ. Electron occupations in terms of alpha ($\upharpoonleft$) and beta ($\downharpoonright$) spins are shown for the neutral ground state S$_0$. 
        (c, d) Dominant electronic configurations of the lowest electronic states D$_0$, D$_1$ and D$_2$ of HYQ$^+$ relevant for PI. Arrows in blue and orange indicate electron excitations calculated in TD-DFT with corresponding transition amplitudes. As the electron configuration of D$_1$ and D$_2$ differ from S$_0$ by more than one electron, CI transition probabilities from S$_0$ to these states will be small so that the PI reaction channel can be expected to be dominated by transitions to D$_0$.}
        \label{fig:FigureS2}
    \end{figure}

The contributions of the different $j$ states to the PI deflection profile of \emph{cis-}HYQ as obtained from the fit are displayed in Figure~\ref{fig:FigureS1} (b). The complete simulated reaction-deflection profile of both \emph{cis-} and \emph{trans-}HYQ is displayed as the black trace in Figure~\ref{fig:Figure2} (c) of the main text where it is also compared with the experimental data. The fit yielded the lowest rotational states, which are most strongly deflected in the experiment, to only negligibly contribute, rationalising the reduced reactivity in PI observed experimentally at the highest deflection coordinates. This can be contrasted with the simulation assuming contributions of the rotational states proportional to their thermal populations in the beam (Figure~\ref{fig:FigureS1} (a)), i.e., assuming that all rotational levels exhibit an equal reactivity. This treatment only yielded an unsatisfactory agreement with experiment (magenta trace in Figure~\ref{fig:Figure2} (c) of the main text).

\pagebreak\clearpage

\subsection*{Molecular orbitals and electronic states}

\pagebreak\clearpage

\subsection*{Geometries of dissociation products}

\begin{figure}[!h]
        \centering
        \includegraphics[width=\textwidth]{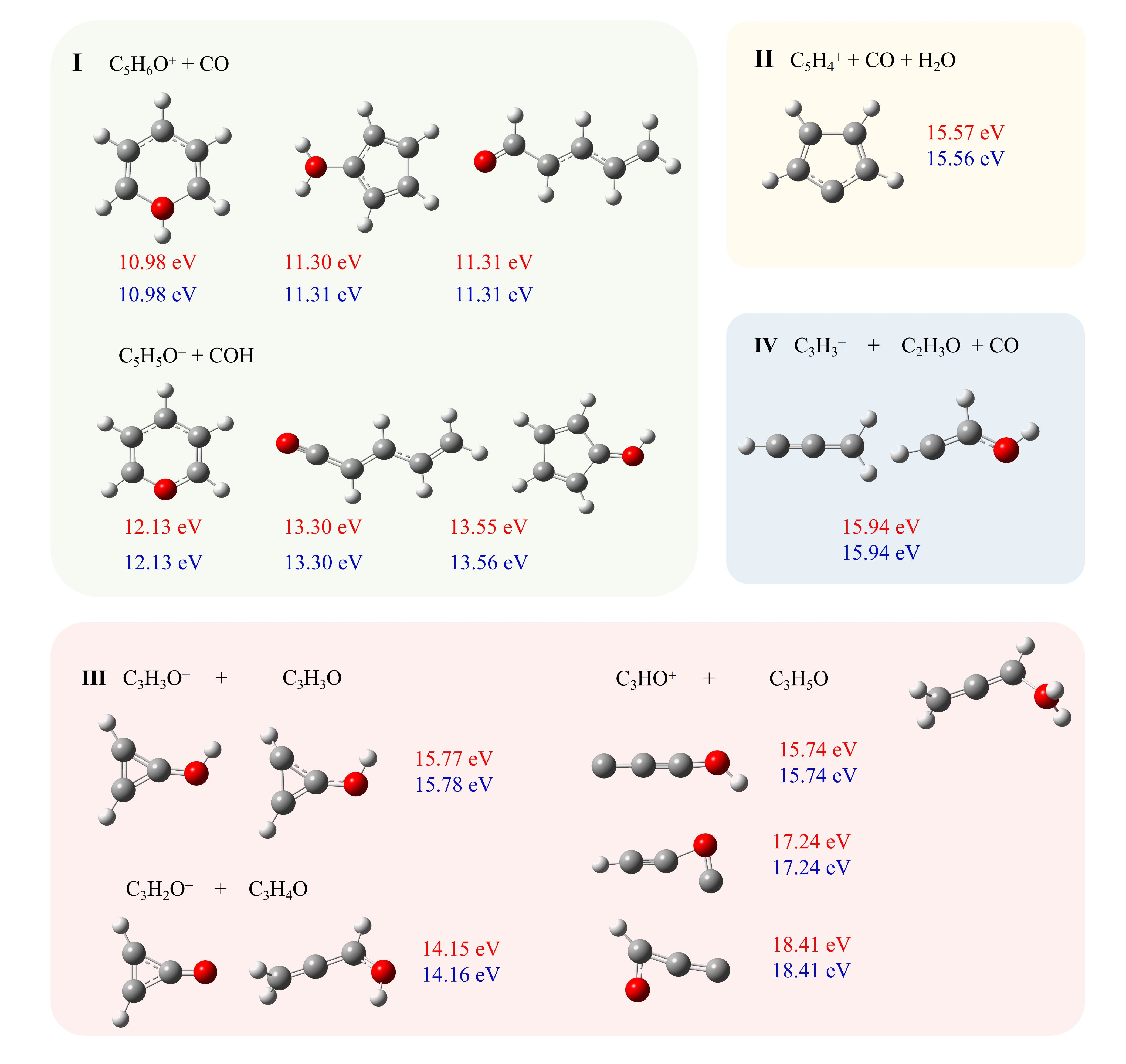}
        \caption{Possible geometries and energies of dissociation asymptotes of DI reactions of HYQ with Ne$^*$. The dissociation energies are indicated in red and blue for \textit{cis-} and \textit{trans-}HYQ, respectively. }
        \label{fig:FigureS3}
    \end{figure}

Figure~\ref{fig:FigureS3} depicts the energies and possible structures of the dissociation products observed for the CI reaction of HYQ with Ne$^*$ listed in Table~\ref{tab:HYQReacProducts} of the main text. Different geometries are accessible for the products of the various various fragmentation channels at the experimentally available energy entailing slightly different dissociation thresholds as indicated. Figure~\ref{fig:Figure3} of the main text displays the relevant channels with the lowest dissociation energies.

\pagebreak\clearpage

\subsection*{Quasiclassical trajectory calculations}

\subsubsection*{Model potential}

\begin{figure}[!h]
        \centering
        \includegraphics[width=1\linewidth]{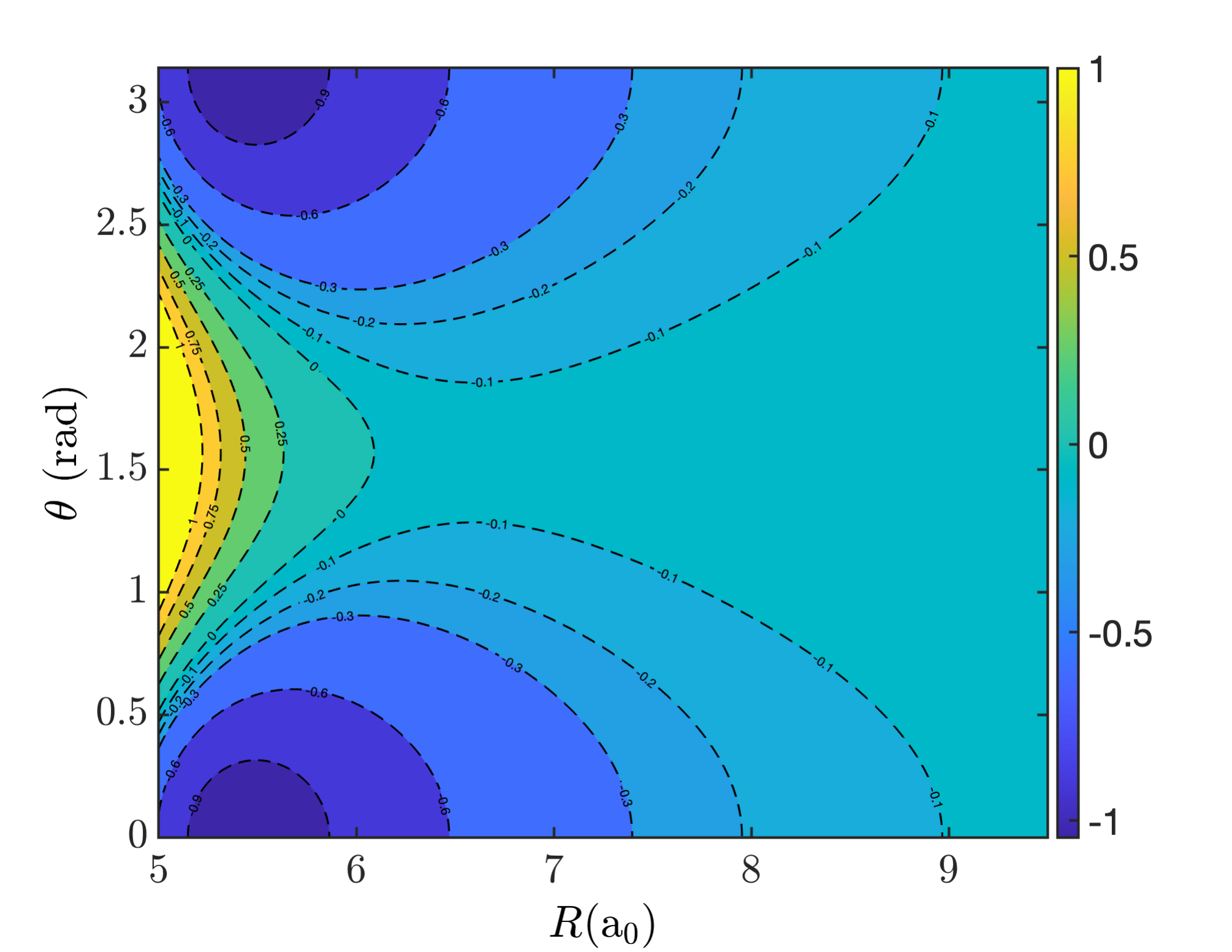}
        \caption{Potential energy landscape due to the dipole-induced dipole interaction in Ne$^*$-HYQ. The colour bar indicates the interaction energy in eV. 
        }
        \label{fig:FigureS4}
    \end{figure}

Figure~\ref{fig:FigureS4} shows a contour plot of the dipole-induced dipole-model potential, including short-range effects, used in the QCT calculations and given by Equation~(\ref{eqqc1}). The potential shows the typical anisotropy of dipole-induced dipole interactions with a a minimum of -0.68~eV at r=5.62~a$_0$ and $\theta=0$.

\subsubsection*{Classical dynamics}

Hamilton's equation are solved via an explicit Runge-Kutta (4,5) approach, as implemented in Matlab. The initial conditions are chosen such that the translational energy (associated with $r$) corresponds to the collision energy and the rotational energy is given by $p_\theta=(j+1/2)/R_e$ (in a.u.). 

\subsubsection*{Solving the time-dependent Schrödinger equation}

\begin{figure}[!h]
        \centering
        \includegraphics[width=1\linewidth]{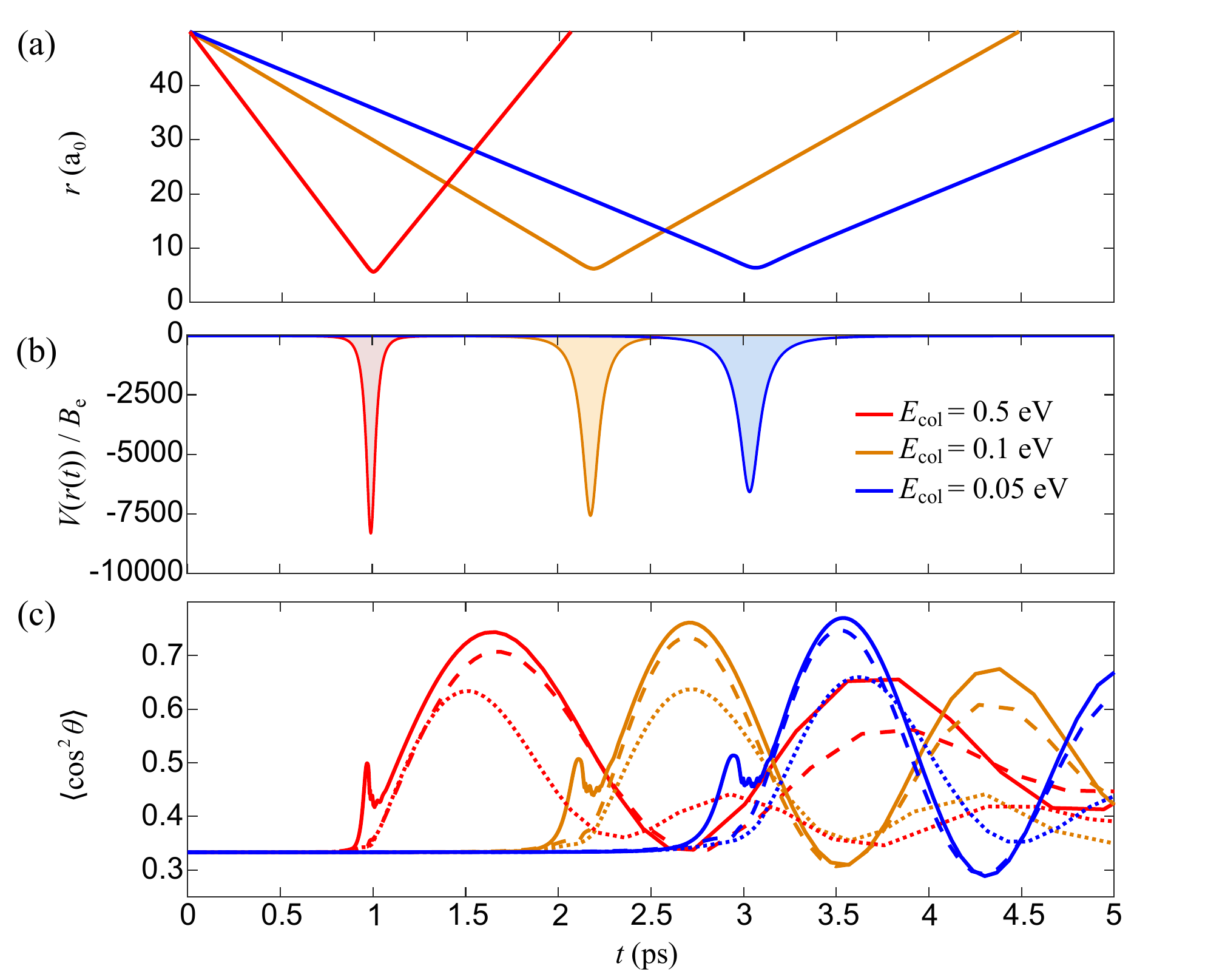}
        \caption{Collision-induced alignment. (a) Atom-molecule distance as a function of time for three different collision energies. Chemi-ionization occurs predominantly around the turning point of the trajectory when orbital overlap between the collision partners is maximal. (b) Time-dependent alignment potential generated by the dipole-induced dipole interaction (Figure~\ref{fig:FigureS4}) during the collision. (c) Time-dependent alignment for three different initial rotational states ($j=0$ (solid line), $j=3$ (dashed line) and $j=10$ (dotted line)) and three collision energies.}
        \label{fig:FigureS5}
    \end{figure}
    
Hamilton's equations and the time-dependent Schrödinger equation (TDSE) were solved simultaneously at each time step of a trajectory, yielding the correct time-dependent interaction potential to describe the internal dynamics of the molecule at every instance of time. The results for three different collision energies are shown in Figure~\ref{fig:FigureS5}. Panel~(a) displays typical trajectories for atom-molecule collisions, giving rise to the time-dependent potentials shown in panel~(b) of the Figure. The time-dependent potential has the shape of a pulse inducing the alignment. The collision energy markedly affects the shape of the time-dependent potential. Panel~(c) displays the time-dependent alignment of the molecule for different initial rotational states. Note the strong dependence of the alignment on the initial rotational state. The lowest rotational states align most easily, whereas rotational excitation suppresses alignment.

The results shown in Figure~\ref{fig:FigureS5} have been calculated via the ansatz of Equation~(\ref{eqqc5}), including rotational states up to $j=100$ for a given value of the orientation quantum number~$m$, which is conserved in the present model. We assumed that the initial $m$-state populations are the same for a given $j$ level, reflecting the loss of any molecular orientation of the molecules on their transit from the electrostatic deflector to the interaction region.

\end{document}